\def\rsol{r_{\odot}}
\def\s12{\sin^2(2\,\theta_{12})}
\def\s13{\sin^2(2\,\theta_{13})}
\def\s23{\sin^2(2\,\theta_{23})}
\def\dmsol{\Delta m^2_\odot}
\def\dmatm{\Delta m^2_{\text{atm}}}
\def\tgatm{\tan^2\theta_{\text{atm}}}
\def\tgsol{\tan^2\theta_\odot}
\def\tgone{\tan^2\theta_{12}}
\def\tgtwo{\tan^2\theta_{23}}
\def\Yl{\textbf{Y}_\ell}
\def\Ynu{\textbf{Y}_\nu}
\def\YU{\textbf{Y}_U}
\def\YD{\textbf{Y}_D}
\def\YR{\textbf{Y}_R}
\def\kmat{\boldsymbol{\kappa}}
\def\Rthe{\textbf{R}_{\,\theta}}
\def\Rphi{\textbf{R}_{\,\phi}}
\def\MR{\textbf{M}_R}
\def\Mnu{{\rm \bf M}_{\nu}^{\rm eff}}
\def \LX{\Lambda_X}
\def \LR{\Lambda_R}
\def \meff{\mbox{$\left|  < \! m  \! > \right| \ $}}
\def\openone{\leavevmode\hbox{\small1\kern-3.3pt\normalsize1}}

\documentclass{JHEP}
\usepackage{amssymb,amsmath,graphicx}

\title{Is right-handed neutrino degeneracy compatible with the solar and
atmospheric neutrino data? }

\author{R. Gonz\'{a}lez Felipe\thanks{E-mail: gonzalez@gtae3.ist.utl.pt}\\
Centro de F\'{\i}sica das Interac\c{c}\~{o}es Fundamentais, Departamento de
F\'{\i}sica,\\ Instituto Superior T\'{e}cnico, 1049-001 Lisboa, Portugal}
\author{F. R. Joaquim\thanks{E-mail: filipe@cfif3.ist.utl.pt}\\
Centro de F\'{\i}sica das Interac\c{c}\~{o}es Fundamentais, Departamento de
F\'{\i}sica,\\ Instituto Superior T\'{e}cnico, 1049-001 Lisboa, Portugal}

\preprint{FISIST/11-2001/CFIF}

\abstract{ In light of the recent solar and atmospheric neutrino
data, we investigate the possibility of having an exactly degenerate
spectrum for heavy right-handed Majorana neutrinos at the grand
unification scale. The analysis is performed in the context of the
minimal supersymmetric standard model with unbroken $R$ parity and
extended with three heavy Majorana neutrino fields in order to
implement the seesaw mechanism. In the absence of a Dirac-type
leptonic mixing, the only source of lepton flavour violation is the
right-handed neutrino sector. Inspired by GUT-motivated relations
among the quark, charged-lepton and Dirac neutrino Yukawa coupling
matrices, and after the inclusion of the radiative effects, we
determine the effective neutrino mass matrix at the electroweak
scale. Using then the latest global analyses of the solar and
atmospheric data at 99\% C.L., we conclude that, within this
framework, the only solar solutions compatible with the experimental
data are the LOW and LMA solutions, being the latter the most
favoured one. At 90\% C.L., only the LMA solution is allowed. }

\keywords{neutrino physics, solar and atmospheric neutrinos, seesaw
mechanism}

\begin{document}

\section{Introduction}

The recent observation of solar and atmospheric neutrino deficits
constitutes a solid evidence for physics beyond the Standard Model.
The existence of such anomalies gives us a strong indication of
neutrino oscillations among the three known families, which in turn
imply that neutrinos have nonvanishing masses and nontrivial mixings.
The latest atmospheric neutrino data from the Super-Kamiokande
Collaboration \cite{SK1} and CHOOZ experiments \cite{CHOOZ} tends to
favour the $\nu_\mu \rightarrow \nu_\tau$ oscillations with a large
mixing angle $\theta_{\rm atm} \simeq \pi/4$. On the other hand, the
solar neutrino experiments \cite{SK2}-\cite{gallex} point towards the
conversions  $\nu_e \rightarrow \nu_\mu$ and $\nu_e \rightarrow
\nu_\tau$, and strongly favour the large mixing angle (LMA) MSW
\cite{MSW} solution over the small mixing angle (SMA), the LOW and
the just-so vacuum oscillation (VO) solutions, as well as over the
sterile neutrino hypothesis.

From a theoretical viewpoint, modelling neutrino masses and mixings
represents a challenge. First, the presently most favoured solution,
namely the LMA MSW solution, is usually the most difficult one to
achieve in realistic models. Second, large mixing angles do not seem
to fit naturally in our grand unification view. This is because in a
very large class of string and grand unified theories (GUT's), the
quark, charged lepton and Dirac neutrino mass matrices are typically
related to each other and, therefore, one would expect small neutrino
mixings in analogy to what is observed in the quark sector. The above
reasoning assumes of course that the large neutrino mixing arises
only from the Dirac neutrino mass matrix structure. Third, even if
large neutrino mixings are consistently incorporated in the general
picture of fermions, a natural question then arises, namely, why
neutrino masses are so tiny compared to the rest of the fermions of
the theory.

A simple and economical way to overcome some of the above
difficulties is to invoke the so-called seesaw mechanism
\cite{seesaw} which, in the presence of heavy right-handed Majorana
neutrinos, $\nu_R\ $, leads to an effective mass matrix for the light
neutrinos. Rather than an assumption, the existence of the $\nu_R$'s
is a natural requirement in the context of all GUT's with a group
symmetry larger than $SU(5)$. The seesaw mechanism by itself has,
unfortunately, a very limited predictive power: the overall scale of
the light neutrino masses is not uniquely determined, the mixing
angles and the ratios of the effective neutrino masses are not fixed.

Grand unified theories such as $SO(10)$ \cite{so10} are a suitable
framework not only to analyze fermion masses but also to implement
the seesaw mechanism. One of the attractive features of the $SO(10)$
model is that its gauge group is left-right symmetric and,
consequently, there exists a complete quark-lepton symmetry in the
spectrum. In particular, the fact that all left-handed (right-handed)
fermions of each family fit into the single irreducible spinor
representation \textbf{16} ($\overline{\textbf{16}}$) of $SO(10)$ and
that the right-handed neutrino $\nu_R$ is precisely contained in this
representation is remarkable. As it happens in all gauge groups,
after the spontaneous symmetry breaking, the fermions will acquire masses
out of the Yukawa couplings and the vacuum expectation values (VEV)
of the Higgs fields. Several constraints on fermion masses are
usually implied in these models \cite{so10}. For instance, if there
is only one \textbf{10} Higgs multiplet responsible for the masses,
then we have the relation $\YU=\YD=\Yl=\Ynu\ $ (the indices $U, D,
\ell$ and $\nu$ stand for the up quarks, down quarks, charged leptons
and Dirac neutrinos, respectively). Similarly, the existence of two
\textbf{10} Higgs multiplets implies $\YD=\Yl$ and $\YU=\Ynu\ $. On
the other hand, if the fermion masses are generated by a VEV of the
\textbf{126} of $SO(10)$, then the $SU(4)$ symmetry yields the
relations $3\, \YD=-\Yl$ and $3\, \YU=-\Ynu\ $. Of course, equalities
such as the ones arising purely from the \textbf{10}-dimensional
representation cannot be exact in realistic models, since they imply
undesirable relations among the quark and charged lepton masses.
Additional assumptions are therefore necessary in order to predict
the correct fermion spectrum \cite{georgi,harvey}.

At this point it is worth recalling that the Majorana mass matrix,
$\MR\ $, for the right-handed neutrinos is usually much less
constrained in unified models than the Dirac mass matrix. This has
led many authors to discuss possible textures for the heavy Majorana
neutrino mass matrix within the seesaw framework
\cite{smirnov}-\cite{altarelli}. A common feature to most of these analyses
is the fact that the heavy Majorana spectrum turns out to be either
hierarchical or partially degenerate. We may therefore ask ourselves
whether a complete degenerate spectrum is allowed for $\MR$ in this
context. Similar studies have been previously performed in the case
of degenerate left-handed Majorana neutrinos
\cite{ellis}-\cite{chun}. In particular, it has recently been argued
\cite{dighe} that if the Majorana masses of the light neutrinos are
exactly degenerate at some high scale, then the one-loop radiative
corrections are not sufficient to produce the required neutrino mass
splittings and mixings at low scale.

Although a fully degenerate mass pattern for the light neutrinos
could be natural, if the Majorana masses arise from nonrenormalizable
operators, and is welcome by cosmology, e.g. if relic neutrinos
constitute the hot dark matter of the Universe, at a fundamental
level it seems more plausible to construct and motivate patterns for
the Dirac and right-handed Majorana matrices rather than for the
effective neutrino mass matrix. In fact, the effective neutrino mass
matrix is only a secondary output of the seesaw mechanism and,
consequently, it may not be directly related to any symmetry of the
theory.

Of course, the observed hierarchy $\Delta m^2_{\rm atm} = |m^2_3 -
m^2_2| \gg \Delta m^2_\odot = |m^2_2 - m^2_1|$ among the light
neutrino masses $m_i\ (i=1,2,3)\ $ implies that one of the following
mass patterns should be realized: $|m_{1,2}| \ll |m_3|$
(hierarchical), $|m_1| \sim |m_2| \gg |m_3|$ (inversely hierarchical)
or $|m_1| \sim |m_2| \sim |m_3|$ (almost degenerate). Thus, if one
believes in the seesaw as the mechanism responsible for the smallness
of neutrino masses, any realistic texture for the Dirac and Majorana
matrices should lead to one of these patterns.

In this paper we address the question whether right-handed neutrino
degeneracy is compatible or not with the present experimental solar
and atmospheric neutrino data. We assume complete degeneracy of the
heavy Majorana neutrinos $\nu_R$ at GUT scale and also that the
right-handed neutrino sector provides all the source of lepton
flavour violation in the theory. Neglecting the rotation between the
charged lepton and the Dirac neutrino mass matrices is a reasonable
approximation, if the Higgs field has its components along suitable
representations of $SO(10)$ such as the ones mentioned above. This
scenario is quite appealing since one starts from a situation where
there is no leptonic mixing (similar to what happens in the quark
sector where the mixing is described by the CKM matrix and is
therefore small) to generate large mixing through the seesaw
mechanism.

When discussing neutrino masses and mixing angles, a crucial aspect
is the running of the various couplings from the unification scale
down to the electroweak scale. In the presence of nonvanishing
neutrino couplings, the renormalization group equations (RGE) are
modified. From the unification scale $\Lambda_X$ to the $\nu_R$
decoupling scale $\Lambda_R$, the radiative corrections from the
Dirac and right-handed Majorana neutrinos must be included. Below
$\Lambda_R$, when the seesaw mechanism becomes operative and the
$\nu_R$'s decouple from the theory, the running of the effective
neutrino Yukawa couplings then becomes relevant. Radiative
corrections could indeed provide potential contributions to the
effective neutrino mass matrix or be even disastrous. A typical
example is given by the RGE effects due to the charged-lepton Yukawa
couplings \cite{RGE}.

The outline of the paper is as follows. In Section \ref{sec2} we
briefly present the general framework used in our analysis, with some
attention given to the parametrization of the degenerate right-handed
Majorana neutrino mass matrix. Section \ref{sec3} is devoted to the
study of the RGE effects from the GUT scale down to the electroweak
scale. In order to account for the right-handed decoupling threshold
at $\Lambda_R$, the RGE analysis consists of two steps: first, the
running of all the quark, charged-lepton as well as the Dirac and
right-handed neutrino Yukawa couplings from $\Lambda_X$ to
$\Lambda_R$ and, second, the evolution of all the quark and
charged-lepton couplings together with the effective neutrino mass
matrix from $\Lambda_R$ to the electroweak scale $M_Z$. Such an
analysis is required in order to verify whether the imposed patterns
for the Yukawa coupling matrices at high scales are compatible or not
with the low-energy data. In Section \ref{sec4}, the effective
neutrino mass matrix, obtained at the electroweak scale, is
confronted with the experimental data to determine which of the
currently allowed solar neutrino solutions can indeed be realized in
the present framework. Some numerical examples for the most favoured
solution, namely the LMA MSW solution, and for different hierarchical
patterns of the Dirac neutrino Yukawa couplings, are given in this
section. Few comments on neutrinoless double $\beta$ decay are also
presented. These results allow us to motivate a simple analytical
structure for the effective neutrino mass matrix in Section
\ref{sec5}. Finally, we present our concluding remarks in Section
\ref{sec6}.

\section{Theoretical framework}
\label{sec2}

We shall work in the context of the minimal supersymmetric standard
model (MSSM) with unbroken $R$ parity and extended with 3 heavy
right-handed neutrinos. Supersymmetry (SUSY) is a natural choice to
implement the seesaw mechanism. For instance, the hierarchy problem,
usually present in a nonsupersymmetric seesaw, is automatically cured
in the supersymmetric case. Indeed, the existence of very heavy
right-handed neutrinos coupled to the Higgs field implies
logarithmically-divergent radiative corrections to the Higgs mass
$\sim \LR^2\Ynu^2H^2\ln(\LR^2/\mu^2)$. While such corrections persist
in nonsupersymmetric theories, they are naturally cancelled in SUSY
by similar contributions coming from the right-handed sneutrinos.

The relevant Yukawa mass terms are described by the Lagrangian
(generation indices are suppressed)
\begin{align}
{\cal L}_{Y} =\ & \bar{Q}\,\YU\,U^c\,H_2+\bar{Q}\,\YD\,D^c\,H_1
+\bar{L}\,\Yl\,E^c\,H_1 \nonumber \\
&+\bar{L}\,\Ynu\,\nu_R^c\,H_2+
\frac{1}{2}\,\nu_{R}^{c\;T} \,\MR\, \nu_R^c+\text{h.c.}\ ,
\label{L}
\end{align}
where $\YU\ , \YD\ , \Yl$ and $\Ynu$ are the up-quark, down-quark,
charged-lepton and Dirac neutrino Yukawa coupling
matrices\footnote{We assume all the Yukawa coupling matrices real.},
respectively; $\MR$ is a $3 {\times} 3$ symmetric Majorana mass matrix which
preserves the SM gauge symmetry; $Q$ and $L$ are the quark and
charged-lepton left-handed doublets; $U^c\ , D^c$ and $E^c$ are the
up-quark, down-quark and charged-lepton right-handed singlets. The
additional neutrino chiral fields $\nu_{R\,i}\ (i=e,\mu,\tau)$ are
right-handed Majorana fields responsible for the seesaw mechanism.
Finally, $H_1$ and $H_2$ are the hypercharge -1/2 and +1/2 MSSM Higgs
doublets, whose neutral components acquire VEVs after the spontaneous
symmetry breaking: $\langle H_1^0 \rangle = v \cos \beta\ ,\ \langle
H_2^0 \rangle = v \sin \beta$, with $v \simeq 174$~GeV.

In what follows we shall assume that $\Ynu \propto \YU$ and also that
there is a left-handed alignment between $\Yl\ $ and $\Ynu$ at the
unification scale $\LX$, which can be equal or higher than the
decoupling scale $\LR\,$, i.e. the scale at which the seesaw
mechanism becomes operative and the $\nu_R$'s decouple from the light
fields. As mentioned in the Introduction, these are reasonable
assumptions in the context of grand unified theories. This means, in
particular, that the only source of non-trivial leptonic mixing is
the right-handed neutrino sector. In this scenario large neutrino
mixings will be generated through the seesaw mechanism.

Next we assume that the right-handed Majorana neutrinos are
degenerate at $\LX$. In this case a simple and general
parametrization \cite{branco} can be given for the right-handed
Majorana neutrino mass matrix $\MR$. Indeed, since $\MR$ is symmetric
we can always write
\begin{align}
W^T\,\MR\,W=\LR \openone\ , \quad \YR \equiv \LR^{-1} \MR =
\,W^{\ast}W^{\dag}\ ,
\end{align}
where $W$ is a unitary matrix and $\YR$ is a unitary symmetric
matrix. It is then straightforward to show that the matrix $\YR$ can
be parametrized as
\begin{align}
\YR=\Rthe\,\Rphi\,{\Rthe}^T\ ,
\label{YR}
\end{align}
where the rotation matrices $\Rthe$ and $\Rphi$ are given by
\begin{align}
\Rthe =\left(\begin{array}{ccc}
 c_\theta    & s_\theta       & 0 \\
 -s_\theta   & \quad c_\theta  \quad  & 0 \\
 0              & 0                 & 1
\end{array}\right)\ ,\quad
\Rphi=\left(\begin{array}{ccc}
 e^{\,i\,\alpha}   & 0              & 0 \\
 0                 & \quad c_{2\phi} \quad    &-s_{2\phi}\\
 0                 & -s_{2\phi}    &-c_{2\phi}
\end{array}\right)\ .
\end{align}
Here we have introduced, for simplicity, the notations $c_\varphi
\equiv \cos \varphi\ , \ s_\varphi \equiv \sin \varphi\ $.

Let us note that the parametrization (\ref{YR}) does not include the
trivial case when $CP$ is conserved and all the right-handed
neutrinos have the same $CP$ parity \cite{branco}. The $CP$-violating
phase $\alpha$ turns out to be irrelevant in our further analysis and
thus we shall set $\alpha=0$. In the latter case the complete form of
$\YR$ is given by
\begin{align}
\YR= \left(\begin{array}{ccc}
c^2_\theta+c_{2\phi} s^2_\theta & -s_{2\theta} s^2_\phi & -s_\theta s_{2\phi} \\
 -s_{2\theta} s^2_\phi & \quad s^2_\theta+c_{2\phi} c^2_\theta \quad
 & -c_\theta s_{2\phi} \\
-s_\theta s_{2\phi} & -c_\theta s_{2\phi}  & -c_{2\phi}
\end{array}\right)\ .
\label{YRf}
\end{align}
Moreover,
\begin{align}
{\MR}^{-1}=\LR^{-1}\,[\,\Rthe\,\Rphi\,{\Rthe}^T\,]^{-1}=
\LR^{-1}\,\Rthe\,\Rphi\,{\Rthe}^T=\LR^{-1}\,\YR\ .
\label{MRinv}
\end{align}
Notice that the parametrization (\ref{YRf}) highly constrains the
form of $\MR$, which only depends on two angles $\theta, \phi$ and
the decoupling scale $\LR\ $.

\section{The effects of radiative corrections}
\label{sec3}

In this section we shall consider the evolution of the couplings from
the unification scale $\LX$ to the electroweak scale $M_Z$. The
relevant one-loop RGE, which include neutrino threshold effects, are
summarized in the Appendix. Let us note that it is customary to take
the SUSY breaking scale as $\Lambda_{\text{SUSY}} \simeq 1$ TeV or
$\Lambda_{\text{SUSY}} \simeq m_t\ $. Since in general the
renormalization group effects that could possibly arise between
$\Lambda_{\text{SUSY}}$ and $M_Z$ are negligible, we will not account
for such effects in our analysis. Thus we set
$\Lambda_{\text{SUSY}}=M_Z$.

At the unification scale $\LX$ the Yukawa matrices $\Yl$ and $\Ynu$
can be diagonalized by real and orthogonal matrices $U_{L,R}$ and
$V_{L,R}$ such that
\begin{align}
&{U_L}^{\dag}\,\Yl\,U_R=\text{diag}\left(\,y_e(\LX)\,,\,y_\mu(\LX)\,,\,
y_\tau(\LX)\,\right)\ , \nonumber \\
&{V_L}^{\dag}\,\Ynu\,V_R=\text{diag}\left(\,y_1(\LX)\,,\,y_2(\LX)\,,\,
y_3(\LX)\,\right)\ , \label{diagYe}
\end{align}
where $y_i(\LX)$ denotes the respective Yukawa couplings at $\LX$.
The hypothesis of left-handed alignment between $\Ynu$ and $\Yl$
implies $U_L=V_L$. In this case, $\Ynu$ and $\Yl$ can be
simultaneously diagonalized, leaving the charged-current Lagrangian
invariant. This means that there is no ``Dirac-type'' leptonic mixing
at $\LX$. Obviously, this diagonalization process does not affect the
spectrum of the right-handed Majorana matrix and, therefore, the
resulting matrix $\MR^{\prime}=V_L^T\,\MR\,V_L$ can also be
parametrized in the form of Eq.~(\ref{YRf}) with a different set of
the parameters $\theta$ and $\phi$.

\subsection{Running $\Ynu$ from $\LX$ to $\LR$}

Our first step in discussing the effects of radiative corrections on
neutrino masses and mixing angles will be the study of the evolution
of the Dirac Yukawa coupling matrix $\Ynu$ from GUT scale to the
right-handed neutrino decoupling scale. After performing the
diagonalization given in Eq.~(\ref{diagYe}) we have
\begin{align}
\Ynu\,(\LX)=y_3(\LX)\, \left(\begin{array}{ccc}
\epsilon_1     & 0              & 0 \\
 0             & \epsilon_2     & 0 \\
 0             & 0              & 1
\end{array}\right)\ ,
\label{YDLX}
\end{align}
where the parameters
\begin{align}
\epsilon_1=\frac{y_1(\LX)}{y_3(\LX)}\ , \quad
\epsilon_2=\frac{y_2(\LX)}{y_3(\LX)}\ ,
\label{e1e2}
\end{align}
have been introduced.

The running of $\Ynu$ from $\LX$ to $\LR$ is governed by
Eq.~(\ref{Ynurge}). From this equation and neglecting all but the
third-generation Yukawa couplings\footnote{This is a reasonable
assumption considering that we will focus on cases where
$y_1 \ll y_2 \ll y_3$.} we get
\begin{align}
\Ynu(\LR)\simeq I_1
\left(\begin{array}{ccc}
1    & 0     & 0 \\
0    & 1     & 0 \\
0    & 0     & I_\nu
\end{array}\right)\Ynu(\LX)\ ,
\label{YDLR1}
\end{align}
where
\begin{align}
I_1
=\exp\left[\frac{1}{16\,\pi^2}
\int_{t_X}^{t_R} (\,T_{\nu}-G_{\nu})\,dt \right]
\simeq \exp \left[\frac{1}{16\,\pi^2}
\int_{t_X}^{t_R} \left(3y_t^2+y_3^2
-\frac{3}{5}g_1^2-3g_2^2\right)\,dt \right],
\label{I1}
\end{align}
\begin{align}
I_\nu  \simeq  \exp\left[\frac{1}{16\,\pi^2\,}
\int_{t_X}^{t_R} (\,3y_3^2+y_{\tau}^2\,)\, dt \right]\ ,
\label{Inu}
\end{align}
with $\ t_X \equiv \ln (\Lambda_X/M_Z)\ $ and $\ t_R \equiv \ln
(\Lambda_R/M_Z)\ $.

From Eqs.~(\ref{YDLX}) and (\ref{YDLR1}) we find
\begin{align}
\Ynu(\LR)=
K_D \left(\begin{array}{ccc}
 \epsilon_1^{\,\prime}   & 0                         & 0 \\
 0                       & \epsilon_2^{\,\prime}     & 0 \\
 0                       & 0                         & 1
\end{array} \right)\ , \quad
 K_D=y_3(\LX)\,I_1\,I_\nu \ , \quad \epsilon_{1,\,2}^{\,\prime}=
 I_\nu^{-1}\,\epsilon_{1,\,2}\ .
\label{eprimes}
\end{align}

\subsection{Running $\MR$ from $\LX$ to $\LR$}

The running of the right-handed Majorana mass matrix $\MR$ from $\LX$
to $\LR$ is described by Eq.~(\ref{MRrge}) given in the Appendix.
This equation allows us to relate the right-handed neutrino mass
matrices $\MR(\LR)$ and $\MR(\LX)$:
\begin{align}
[\MR(\LR)]_{ij}=[\MR(\LX)]_{ij}\,\exp\left[\,\frac{1}{8\,\pi^2}
\int_{t_X}^{t_R} \left(y_i^2+y_j^2\right)\,dt \right]
\label{YRLR}\ .
\end{align}
In the limit where only the third Dirac neutrino Yukawa coupling
$y_3$ is kept, we find
\begin{align}
\MR(\LR)=\left(\begin{array}{ccc}
 1     & 0     & 0 \\
 0     & 1     & 0 \\
 0     & 0     & I_R
\end{array} \right)\MR(\LX)
\left(\begin{array}{ccc}
 1     & 0     & 0 \\
 0     & 1     & 0 \\
 0     & 0     & I_R
\end{array}\right)\ ,
\label{finalMR}
\end{align}
where
\begin{align}
I_R \simeq \exp\left[\frac{1}{8\,\pi^2}
\int_{t_X}^{t_R} y_3^2\,dt \right]\ .
\label{IR}
\end{align}
The inverse of $M_R$ at $\LR$ is then given by
\begin{align}
\MR^{-1}(\LR)=\LR^{-1}\,\left(\begin{array}{ccc}
 1     & 0     & 0 \\
 0     & 1     & 0 \\
 0     & 0     & I_R^{-1}
\end{array} \right)\YR
\left(\begin{array}{ccc}
 1  & 0   & 0 \\
 0  & 1   & 0 \\
 0  & 0   & I_R^{-1}
\end{array}\right)
\label{fMRinv}\ ,
\end{align}
where $\YR$ has the form given in Eq.~(\ref{YRf}).

\subsection{$\Mnu$ from high to low scales}

In order to study the low-energy consequences of our assumptions,
imposed at high scales, we must find the effective neutrino mass
matrix $\Mnu$ at the typical neutrino experiment scale, i.e. the
electroweak scale ($\simeq M_Z$). If $R$ parity is an exact symmetry
in the MSSM, the lowest (5-dimensional) operator in the
superpotential which violates lepton number and generates a Majorana
mass for the left-handed neutrinos, $\nu$, is given by
\cite{weinberg,babu}
\begin{align}
W_{\nu\nu}=\frac{1}{4}\kmat\,\nu^T\,\nu\,H_2^0\,H_2^0 + \text{h.c.}\ ,
\label{W5}
\end{align}
where $\kmat$ is a $3 {\times} 3$ symmetric matrix. The effective neutrino
mass matrix is obviously given by
\begin{align}
\Mnu=\kmat \langle H_2^0 \rangle^2=\kmat\,v^2\sin^2\beta \ .
\label{Meff}
\end{align}

The supersymmetric seesaw mechanism is realized by adding
to the superpotential the following terms:
\begin{align}
W=W_{\text{MSSM}}-\frac{1}{2}{{\nu}_R^c}^T\,\MR\,{\nu}_R^c+{{\nu}_R^c}^T
\Ynu\,L\,H_2\ . \label{WMNR}
\end{align}
Below $\LR$ one can consider the effective superpotential given
by
\begin{align}
W_{\text{eff}}=
W_{\text{MSSM}}+\frac{1}{2}{(\Ynu\,L\,H_2)}^T\,\MR^{-1}\,(\Ynu\,L\,H_2)\ ,
\label{Weff}
\end{align}
which is obtained by integrating out the heavy right-handed neutrino
fields $\nu_R$ in Eq.~(\ref{WMNR}). The second term in $W_{\text{eff}}$
will originate in the Lagrangian a Majorana mass term of the type
\begin{align}
{\cal L}_{\nu\nu}=-\frac{1}{2}\,\nu^T\kmat\,\nu\,H_2^{\,0}\,H_2^{\,0}
+\text{h.c.}\ ,
\label{LMaj}
\end{align}
where
\begin{align}
\kmat=\Ynu^T\,\MR^{-1}\,\Ynu\,. \label{seesaw}
\end{align}

Let us note that at this stage the matrices $\kmat$ and $\Mnu$ are
defined at the decoupling scale $\LR$. To compare them with the
experimental ones, obtained at the electroweak scale, we must run
these matrices down to $M_Z$ through the corresponding RGE. Taking
into account Eqs.~(\ref{eprimes}) and (\ref{fMRinv}) we have
\begin{align}
\kmat(\LR)=\LR^{-1}\,\Ynu^T(\LR)\,\YR\,\Ynu(\LR) =
\frac{K_D^2}{\LR}\left(\begin{array}{ccc}
\epsilon_1^{\,\prime}      & 0                        & 0 \\
 0                         & \epsilon_2^{\,\prime}    & 0 \\
 0                         & 0                        & I_R^{-1}
\end{array}\right)\,\YR\,\left(\begin{array}{ccc}
\epsilon_1^{\,\prime}      & 0                        & 0 \\
 0                         & \epsilon_2^{\,\prime}    & 0 \\
 0                         & 0                        & I_R^{-1}
\end{array}\right)\ .
\label{KMR2}
\end{align}

The running of the matrix $\kmat$ from $\LR$ to $M_Z$ is described by
the RGE given in Eq.~(\ref{krge}). From the latter equation, and
neglecting the charged-lepton Yukawa couplings $y_{\mu}$ and $y_e$,
we get the following relation between $\kmat(M_Z)$ and $\kmat(\LR)$:
\begin{align}
\kmat(M_Z)=I_2\left(\begin{array}{ccc}
1     & 0     & 0 \\
0     & 1     & 0 \\
0     & 0     &I_{\tau}
\end{array}\right)\,\kmat(\LR)\,\left(\begin{array}{ccc}
1     & 0     & 0 \\
0     & 1     & 0 \\
0     & 0     &I_{\tau}
\end{array} \right)\ ,
\label{kMZ4}
\end{align}
where
\begin{align}
I_2 = \exp\left[\frac{1}{16\,\pi^2}\,
\int_{t_R}^{t_Z} (\,T_{\kmat}-G_{\kmat})\,dt \right]
\simeq\exp\left[\frac{1}{16\,\pi^2}\,\int_{t_R}^{t_Z} \left(6y_t^2
-\frac{6}{5}g_1^2-6g_2^2\right)\,dt \right]\ ,
\label{I2}
\end{align}
\begin{align}
I_{\tau}=\exp\left[\frac{1}{16\,\pi^2\,}\int_{t_R}^{t_Z}
y_{\tau}^2\,dt \right]\ .
\label{Itau}
\end{align}
Using Eq.~(\ref{KMR2}) we finally obtain
\begin{align}
\kmat(M_Z)=\frac{K_{eff}}{\LR} \left(\begin{array}{ccc}
\delta_1     & 0              & 0 \\
 0           & \delta_2       & 0 \\
 0           & 0              & 1
\end{array}\right)\,\YR(\LX)\,\left(\begin{array}{ccc}
\delta_1     & 0              & 0 \\
 0           & \delta_2       & 0 \\
 0           & 0              & 1
\end{array}\right)\ ,
\label{kMZ6}
\end{align}
with
\begin{align}
\delta_{1,\,2}&=I_{eff}\,I^{-1}_{\tau}\,\epsilon_{1,\,2}\ , \quad
I_{eff} = I_R\,I_\nu^{-1}\ ,
\label{delta12} \\ \nonumber \\
K_{eff}&=y_3^2(\LX)\,I_1^2\,I_2\,I_{\tau}^2\,I_{eff}^{-2}\ ,
\label{Kpar}
\end{align}
and $I_1\ ,\ I_2\ $ given in Eqs.~(\ref{I1}) and (\ref{I2}),
respectively.

From Eqs.~(\ref{YRf}) and (\ref{kMZ6}) we obtain our final expression
for the neutrino effective operator at the electroweak scale:

\begin{align}
\kmat(M_Z)=\frac{K_{eff}}{\LR} \left(\begin{array}{ccc}
\delta_1^2\,(c^2_\theta+c_{2\phi} s^2_\theta)
& -\delta_1\,\delta_2\,s_{2\theta} s^2_\phi
& -\delta_1\,s_\theta s_{2\phi} \\
-\delta_1\,\delta_2\,s_{2\theta} s^2_\phi
& \quad \delta_2^2\,(s^2_\theta+c_{2\phi} c^2_\theta) \quad
& -\delta_2\,c_\theta s_{2\phi} \\
-\delta_1\,s_\theta s_{2\phi}
& -\delta_2\,c_\theta s_{2\phi}
& -c_{2\phi}
\end{array}\right)\ .
\label{KMZfin}
\end{align}

Let us remark that the mass spectrum and mixings arising from this
matrix are invariant under the transformations ($\theta \rightarrow
\theta\ ,\ \phi \rightarrow \pi - \phi$) and ($\theta \rightarrow
\pi-\theta\ ,\ \phi \rightarrow \phi$), and, obviously, under any
transformation resulting from their successive application. This
simply corresponds to a phase redefinition of the left-handed
neutrino fields. Thus, from now on we restrict ourselves to the
intervals $0 \leq \theta,\phi \leq \pi/2$.

In the limit where the evolution of the third Dirac neutrino Yukawa
coupling $y_3$ and the $\tau$ Yukawa coupling $y_\tau$ can be
neglected, we find the approximate expressions:
\begin{align}
I_\nu\simeq\left({\frac{\LR}{\LX}}\right)^{\frac{3y_3^2+y_\tau^2}{16\pi^2}}\ ,
\quad
I_R \simeq\left({\frac{\LR}{\LX}}\right)^{\frac{y_3^2}{8\pi^2}}\ ,
\quad
I_{\tau}\simeq\left(\frac{M_Z}{\LR}\right)^{\frac{y_{\tau}^2}{16\pi^2}}\ ,
\label{Iapp}
\end{align}
and, consequently,
\begin{align}
\delta_{1,\,2} &\simeq
\left({\frac{\LX}{\LR}}\right)^{\frac{y_3^2+y_\tau^2}{16\pi^2}}
\left(\frac{\LR}{M_Z}\right)^{\frac{y_{\tau}^2}{16\pi^2}}\,\epsilon_{1,\,2}\ ,
\label{delta12app} \\
K_{eff} &\simeq
y_3^2(\LX)\,I_1^2\,I_2\,\left({\frac{\LR}{\LX}}\right)^{\frac{y_3^2+y_\tau^2}{8\pi^2}}
\left(\frac{M_Z}{\LR}\right)^{\frac{y_{\tau}^2}{8\pi^2}}\ .
\label{Keffapp}
\end{align}

At this point few remarks are in order. First, as it has already been
noticed by several authors \cite{RGE}, the parameter $I_\tau \simeq
1$. This can be easily checked by using either the exact expression
(\ref{Itau}) or the approximate expression given in Eq.~(\ref{Iapp}).
Moreover, the parameter $I_{eff}$, which controls the RGE effects of
the Dirac neutrino Yukawa coupling ratios (cf. Eq.~(\ref{delta12})),
is also close to the unity. This can be seen from Fig.~\ref{fig1}
where we present the dependence of $I_{eff}$ on the value of the
scale $\mu$ for the two initial conditions $\Ynu = \YU$ and $\Ynu =
3\,\YU$. We note that $I_{eff} \simeq 1$ for a wide range of $\tan
\beta$. Notice also that for the case $\Ynu = 3\,\YU$, the use of the
approximate expression for $I_{eff}$, obtained by neglecting the
running of the Yukawa couplings, is not well justified. This is
because in the latter case $y_3(\LX) = 3 y_t(\LX)$ and the evolution
of the couplings becomes more relevant.

\begin{center}
\FIGURE[!ht]{ \label{fig1} \caption{The evolution of the RGE
coefficient $I_{eff} = I_R\,I_\nu^{-1}$ from $\LX$ to $\LR\ $,
assuming two different Dirac neutrino textures, $\YD=\YU$ and $\YD=3
\YU\ $, and for the values of $\tan \beta=10$ and $\tan \beta=55$.
The solid curves correspond to the exact expressions as defined by
Eqs.~(\ref{Inu}) and (\ref{IR}). The dashed lines correspond to the
approximate curves as given in Eqs.~(\ref{Iapp}).}
\begin{tabular}{cc}
\includegraphics[width=7.cm]{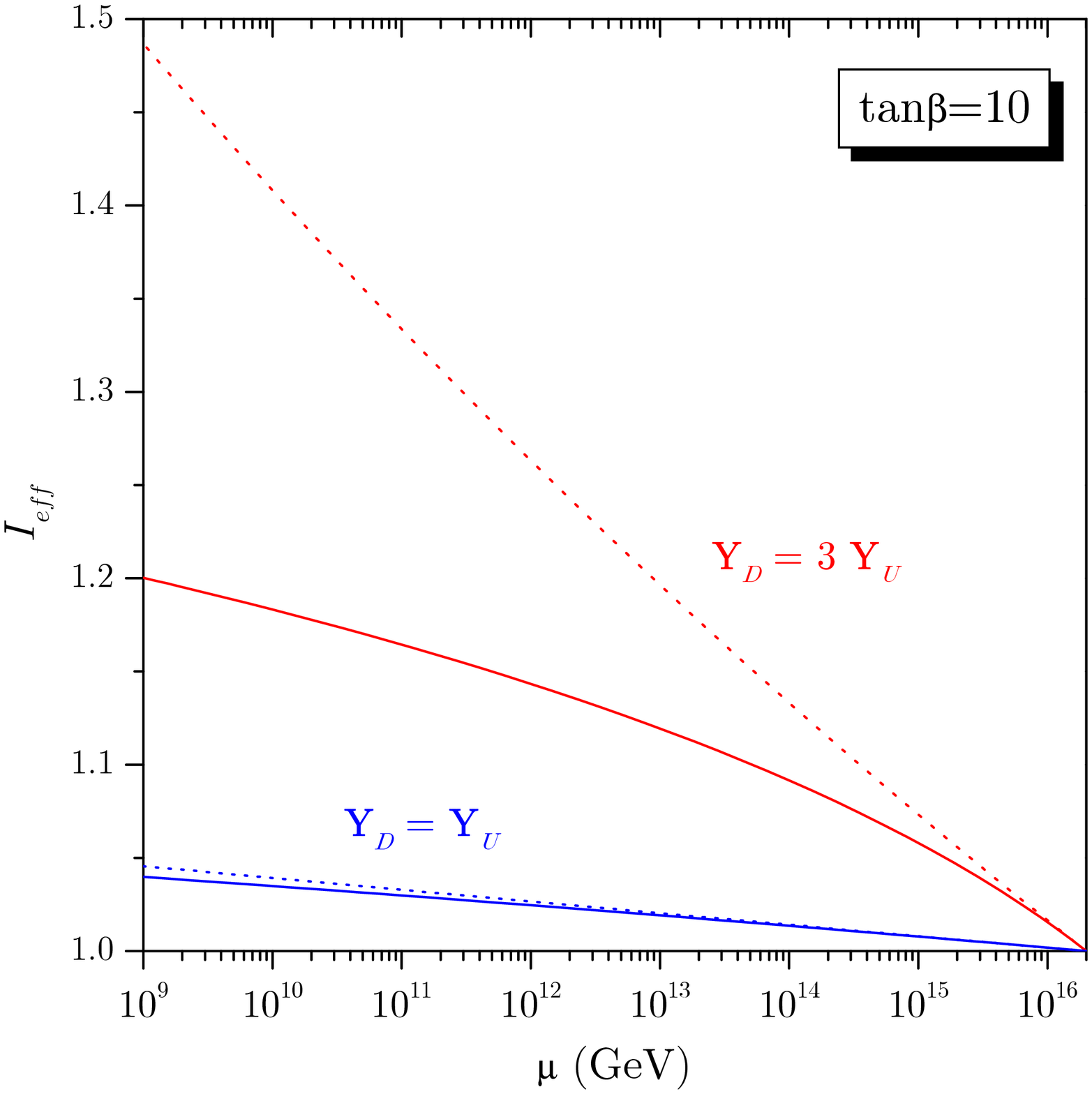}
&\includegraphics[width=7.cm]{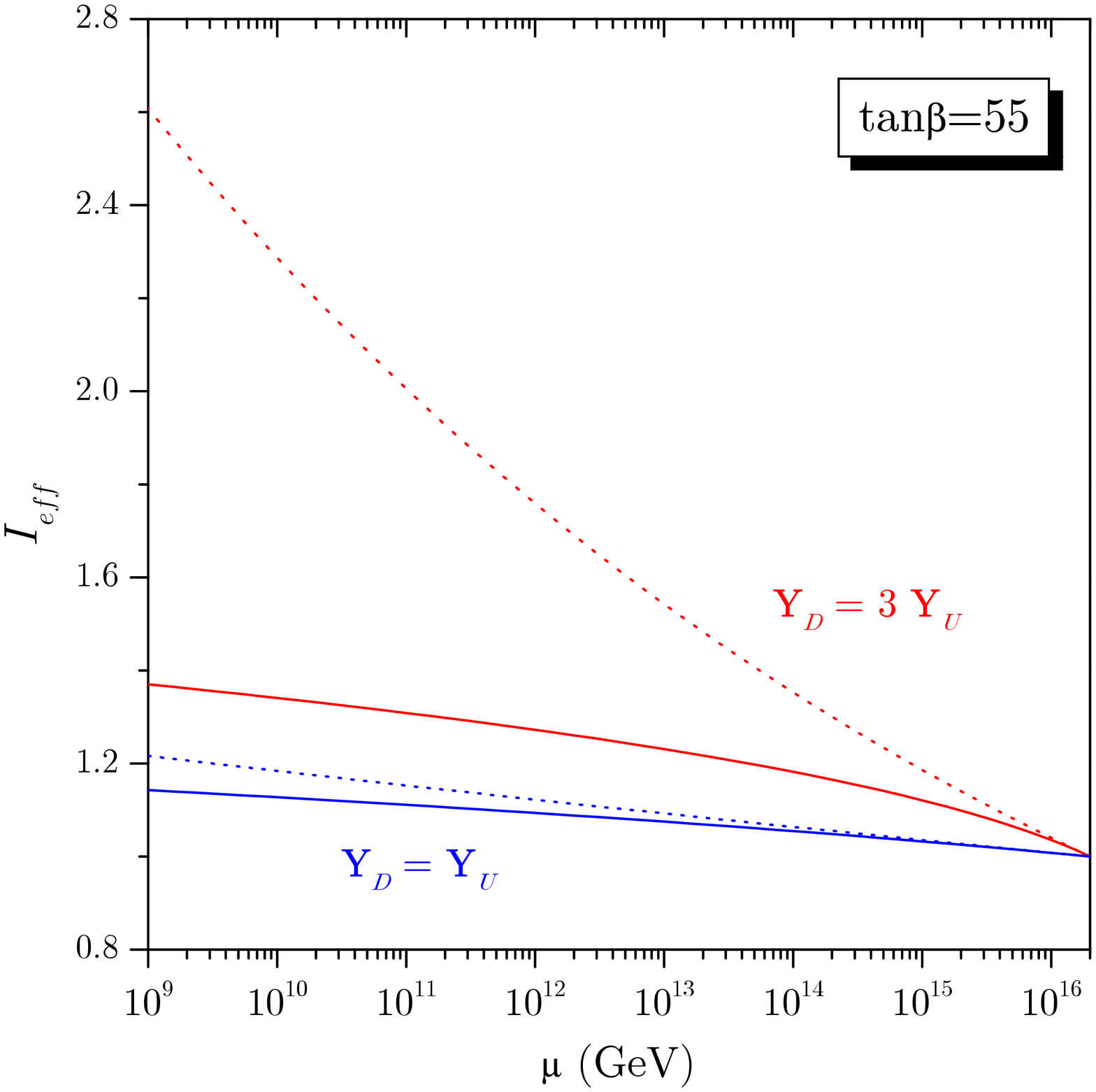}\\
\end{tabular}
}
\end{center}

\section{Confronting the neutrino mass matrix with the experimental
data}
\label{sec4}

The analysis of the neutrino oscillation data indicates that,
contrary to what happens in the quark sector, where the mixing is
always small, the mixing angle is large in the 2-3 neutrino sector.
The latest atmospheric neutrino data from the Super-Kamiokande
Collaboration \cite{SK1} and CHOOZ experiments \cite{CHOOZ} implies a
large mixing angle $\theta_{\text{atm}} \simeq \pi/4$.  In what
concerns the 1-2 neutrino sector, there are four different scenarios
depending on the solar neutrino problem solution.

Since the present experimental data constrains the neutrino mass
squared differences and the mixing angles, it is necessary to
diagonalize the effective neutrino mass matrix $\Mnu$ at the
electroweak scale. The neutrino mixing matrix $\textbf{U}$, which
relates the flavour and mass eigenstates,
\begin{align}
\left( \begin{array}{c}
\nu_e \\ \nu_\mu \\ \nu_\tau
\end{array} \right) = \textbf{U}
\left( \begin{array}{c}
\nu_1 \\ \nu_2 \\ \nu_3
\end{array} \right)\ ,
\end{align}
can be parametrized in the standard form \cite{PDG}
\begin{align}
\textbf{U} =\left(
\begin{array}{ccc}
c_{13} c_{12} & s_{12} c_{13} & s_{13} \\
-s_{12} c_{23} - s_{23} s_{13} c_{12}
& \quad c_{23} c_{12} - s_{23} s_{13} s_{12} \quad
& s_{23} c_{13} \\
s_{23} s_{12} - s_{13} c_{23} c_{12}
& -s_{23} c_{12} - s_{13} s_{12} c_{23}
& c_{23} c_{13}
\end{array}\right) \ ,
\label{UMNS}
\end{align}
such that $\textbf{U}^T\,\Mnu(M_Z)\,\textbf{U} = \text{diag}(\,m_1\ ,
m_2\ , m_3\,)$, where $m_i$ are the light neutrino masses; $c_{ij}
\equiv \cos \theta_{ij}\ , \ s_{ij} \equiv \sin \theta_{ij}\ $.

We also remark that the lepton mixing matrix can have, besides a
Dirac-type phase (analogous to that of the quark sector), two other
physical phases associated with the Majorana character of neutrinos.
If $CP$ is conserved these phases are either 0 or $\pi$. In the
analysis that follows we shall assume all these phases vanishing.
However, as we shall see in Section~\ref{sec4-2}, such phases play a
significant role in processes such as the neutrinoless double $\beta$
decay \cite{bilenky}.

\subsection{Neutrino oscillation data}
\label{sec4-1}

The relevant parameter set for the study of solar and atmospheric
data is usually identified with the quantities
\begin{align}
&\Delta m^2_{\odot} \equiv  \Delta m^2_{12} = m^2_2 - m^2_1\ , \quad
\Delta m^2_{\text{atm}} \equiv \Delta m^2_{23} = m^2_3 - m^2_2\ , \nonumber \\
&\tan^2 \theta_{\odot}  \equiv \tan^2\theta_{12}\ ,\quad
\tan^2\theta_{\text{atm}} \equiv \tan^2\theta_{23}\ , \quad
U_{e3} \equiv \sin \theta_{13}\ ,
\label{paramset}
\end{align}
where all mixing angles are assumed to lie in the range $[0,\pi/2]$.
In Table~\ref{tab1} we summarize the present constraints on neutrino
mass squared differences and on the mixing angles obtained from the
recent global analyses \cite{gonzalez,bahcall} of the solar,
atmospheric and reactor neutrino data at 99\% C.L. [90\% C.L.] .

\TABLE[!ht]{ \label{tab1} \caption{Constraints on neutrino masses and
mixing angles at 99\% C.L. [90\% C.L.] coming from global analyses of
the solar, atmospheric and reactor neutrino data
\protect\cite{gonzalez,bahcall}. The numbers in bold correspond to the best-fit
values.}
\renewcommand{\tabcolsep}{1.06pc}
\begin{tabular}{lccc}
\hline \hline
\noalign{
\begin{center}
Atmospheric and reactor neutrinos
\end{center}
}
\hline \noalign{\smallskip}
 & $\dmatm~(\text{eV}^2)$ & $\tgatm$ & $|U_{e3}|$ \\
\noalign{\smallskip} \hline \noalign{\smallskip}
& $(\,1.1 - 7.3 \,) {\times} 10^{-3}$ & 0.3 - 3.8  & $ < 0.27$ \\
& $[\,1.4 - 6.1 \,] {\times} 10^{-3}$ & [\,0.4 - 3.1\,] & $ [\,< 0.23\,]$ \\
& $ \mathbf{3.1 \times 10^{-3}} $ & \textbf{1.4} & \textbf{0.07} \\
\noalign{\medskip} \hline \hline
\noalign{
\begin{center}
Solar neutrinos
\end{center}
}
\hline \noalign{\smallskip} \noalign{\smallskip}
 & $\dmsol~(\text{eV}^2)$ & $\tgsol$ & $\rsol=\dmsol/\dmatm$ \\
\noalign{\smallskip} \hline \noalign{\smallskip}
LMA & $(\,2.0 - 20 \,) {\times} 10^{-5} $
& 0.1 - 0.7  & $(\,0.03 - 1.8\,) {\times} 10^{-1} $ \\
& $[\,2.3 - 10 \,] {\times} 10^{-5} $
& [\,0.2 - 0.5\,]  & $[\,0.04 - 0.7\,] {\times} 10^{-1} $ \\
& $ \mathbf{4.2 \times 10^{-5}} $ & \textbf{0.26} & \textbf{0.014} \\
\noalign{\smallskip}
SMA & $(\,0.4-1\,) {\times} 10^{-5}$ & $(\,0.2-2\,) {\times} 10^{-3}$
& $(\,0.55-9.1\,) {\times} 10^{-3}$ \\
\noalign{\smallskip}
LOW & $(\,0.5-2\,) {\times} 10^{-7}$  & $0.5-0.9$
& $(\,0.07-1.8\,) {\times} 10^{-4}$ \\
\noalign{\smallskip}
Just-So$^2$ & $(\,5-8\,) {\times} 10^{-12}$  & $0.5-2$
& $(\,0.68-7.3\,) {\times} 10^{-9}$ \\
\noalign{\smallskip} \hline \hline
\end{tabular}
}

Apart from the overall factor $K_{eff}/\LR\,$, the effective neutrino
operator $\kmat(M_Z)$ defined in Eq.(\ref{KMZfin}) depends on the
angles $\theta$ and $\phi$, which parametrize the heavy right-handed
neutrinos, and the parameters $\delta_{1,\,2}$. For the structure of
$\kmat$ to be compatible with the low-energy data, we should take
into account not only all the constraints on $\rsol=\dmsol/\dmatm$
and the mixing angles, but also the experimentally allowed region for
the up-quark mass ratios, since within our assumptions, these ratios
fix the parameters $\delta_1$ and $\delta_2$ at $\LX$:
\begin{align}
\delta_1 \simeq \epsilon_1 = \frac{y_u(\LX)}{y_t(\LX)}\ , \quad
\delta_2 \simeq \epsilon_2 = \frac{y_c(\LX)}{y_t(\LX)}\ .
\end{align}

The above ratios can be determined by running the RGE from the
electroweak scale $M_Z$ to the unification scale $\LX$, which we
assume to be $\LX \simeq 2 {\times} 10^{16}$. Since these ratios are quite
insensitive to the value of $\tan \beta$ in a wide region, we shall
take $\tan \beta = 10$ as the input value in our analysis. Using the
experimentally allowed values for the quark masses at the electroweak
scale \cite{koide, PDG}, the following ranges are then obtained at
$\LX$:
\begin{align}
2.8 {\times} 10^{-6} &\lesssim \delta_1 \lesssim 1.5 {\times} 10^{-5}\ , \nonumber \\
9 {\times} 10^{-4} &\lesssim \delta_2 \lesssim 4 {\times} 10^{-3}\ .
\label{upratios}
\end{align}

In our numerical study we randomly varied all the parameters in the
ranges: $0 \leq \delta_{1,\,2}\leq 1\ ,\ 0 \leq \theta \leq \pi/2\ ,\
0 \leq \phi \leq \pi/2\ .$ In Fig.~\ref{fig2} we plot the allowed
region in the plane ($\delta_1,\delta_2$) for the three different MSW
solar solutions (LMA, SMA, LOW) at 99\% C.L.~. We assume that the
right-handed Majorana neutrinos are exactly degenerate at the
unification scale $\LX$. The dashed area delimited by the dotted
lines corresponds to the allowed region for the Dirac neutrino
coupling ratios, as defined by Eqs.~(\ref{upratios}) and assuming
$\Ynu \propto \YU$ at $\LX$. The intersection of the filled areas
with the dashed area provides us with the points
($\delta_1,\delta_2$) compatible with the low-energy data and the
initial conditions imposed at $\LX$. From the plots it is clear that
the SMA solution is excluded. The LOW solution is barely compatible
with the data at 99\% C.L., and certainly excluded at 90\% C.L.~.
Finally, the LMA solution is the most favoured one, and at 90\% C.L.
it is the only one that survives.

The results in the parameter space ($\phi,\theta$) are shown in Fig.
\ref{fig2}d at 99\% C.L.\,. We notice that the allowed region for LMA
is more extensive than the SMA and LOW regions, and the SMA region is
in turn wider than the LOW one. This is a direct consequence of the
hierarchy ${\rsol}^{\text{LMA}} > {\rsol}^{\text{SMA}} >
{\rsol}^{\text{LOW}}$. For the VO (Just-So$^2$) solar solutions an
extreme fine-tuning is required between $\theta$ and $\phi$ in order
to get a solution compatible with the solar and atmospheric neutrino
data. The latter solutions are therefore not presented in our plots.

Since the LMA solution is the most favoured in the present framework,
the rest of our discussion will be devoted to it. In Figs.~\ref{fig3}
we show the allowed region at 99\% C.L. for the LMA MSW solution, in
different planes of the input parameters $\delta_1,\delta_2, \phi$
and $\theta$. A remarkable feature is the fact that for small values
of $\delta_1$ and $\delta_2$, which are required in our framework,
one has $\phi \simeq \pi/4$ and $\theta \simeq \pi/2$. We remark that
the same conclusion for the value of $\phi$ can be drawn by looking
at the structure of the effective neutrino mass matrix
(\ref{KMZfin}), since for small values of $\delta_{1,\,2}\ $, the
relation $\cos 2\phi \ll 1$ is required to avoid a quasi-diagonal
pattern in the mass matrix.

Let us conclude this section by giving few numerical examples (see
Table~\ref{tab2}). We consider two different GUT-motivated relations
for the Dirac neutrino Yukawa coupling matrix: $\Ynu = \YU$ and $\Ynu
= 3 \YU$. In both cases, consistency with the present LMA solar,
atmospheric and reactor neutrino data was achieved. It was not
possible however to obtain the best-fit values, simultaneously for
all the parameters. By slightly modifying the initial assumption
$y_1/y_3 = y_u/y_t\ $, we were then able to reproduce the best-fit
values for all the solar and atmospheric neutrino data, as shown in the
last column of the table. Notice also that the obtained value
$|U_{e3}| \simeq 0.05$ is close to the best-fit value
$|U_{e3}| = 0.07$ given in Table~\ref{tab1}.

As suggested by our full numerical analysis (see Figs.~\ref{fig3}),
the angles $\theta$ and $\phi$ were chosen to be close to $\pi/2$ and
$\pi/4$, respectively. The degeneracy imposed at $\LX$ for the
right-handed Majorana neutrinos, $|M_1| = |M_2| = |M_3|\ $, is lifted
in all cases by radiative corrections. At the decoupling scale $\LR$
we have $|M_1| \simeq |M_2| \simeq I_R \LR < |M_3| \simeq \LR$. For
the effective neutrinos with masses $m_i\ $, we obtain at the
electroweak scale a hierarchical spectrum with
\begin{align}
|m_1| < |m_2| \simeq \sqrt{\Delta m^2_{12}} < |m_3| \simeq
\sqrt{\Delta m^2_{23}}\ .
\end{align}
Finally, we also notice the stability of the atmospheric and solar
neutrino parameters, $r_\nu = \Delta m^2_{12}/\Delta m^2_{23}\ ,
\ \tgone\ ,\ \tgtwo$ and $U_{e3}$ with the energy scale. In fact,
from $\LR$ to $M_Z$ no significant running of these parameters
is observed. This is related with the fact that $I_\tau \simeq 1$.

\FIGURE[!ht]{ \label{fig2} \caption{Allowed region in the plane
($\delta_1,\delta_2$) for the three different MSW solar solutions
(LMA, SMA, LOW) at 99\% C.L., if we assume $\Ynu \propto \YU$ and an
exactly degenerate spectrum for the right-handed Majorana neutrinos
at the unification scale $\LX$. The dashed area corresponds to the
allowed region for the up quark mass ratios at $\LX$. At 90\% C.L.,
the only solution that is compatible with the experimental data is
the LMA one. The corresponding allowed region in the plane
($\phi,\theta$) is shown in Fig. \ref{fig2}d at 99\% C.L.\,.}
$$\includegraphics[width=14cm]{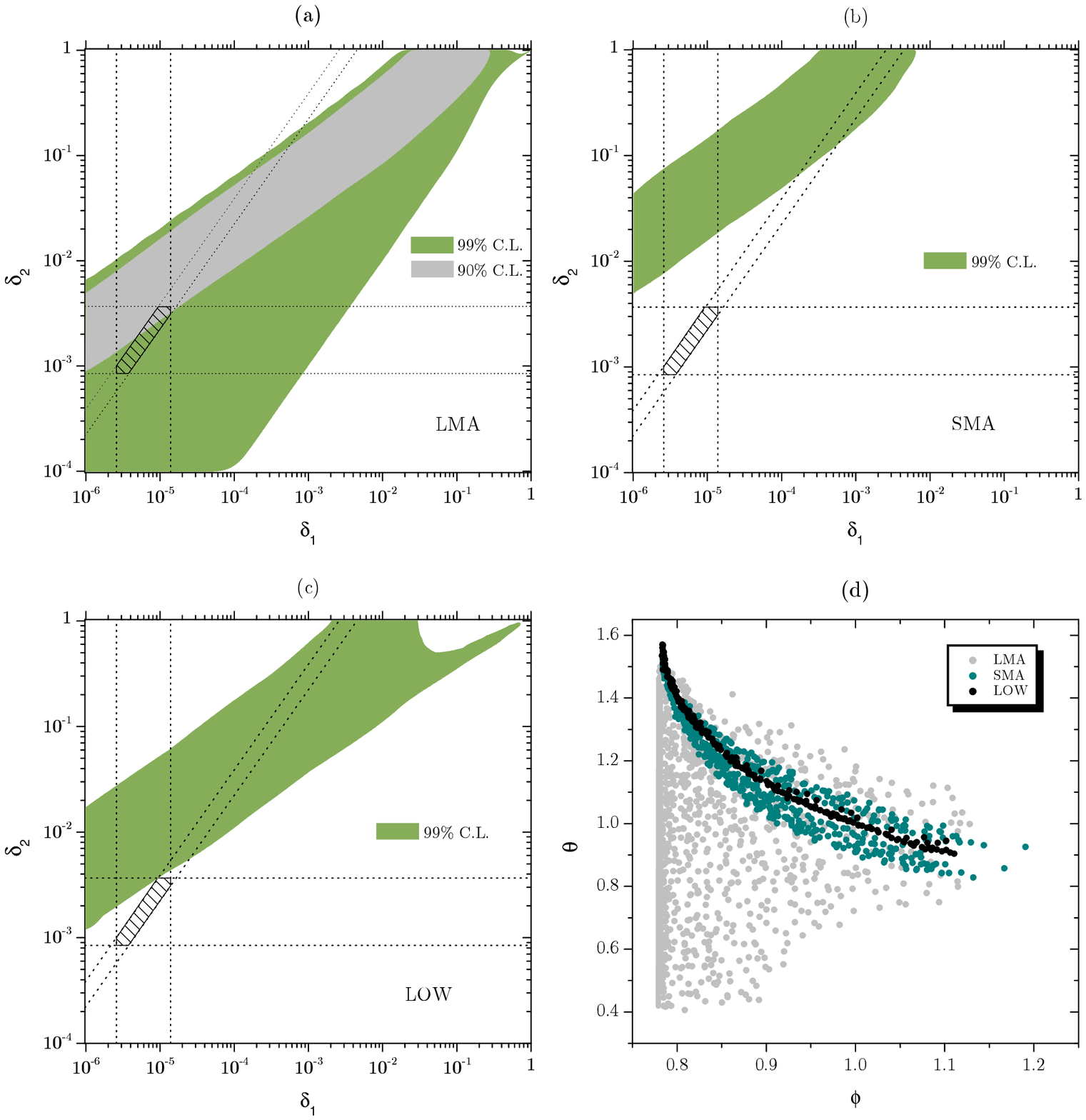}$$
}

\FIGURE[!ht]{ \label{fig3} \caption{Allowed region at 99\% C.L. for
the most favoured solar solution, namely the LMA MSW solution, in
different planes of the input parameters $\delta_1,\delta_2, \phi$
and $\theta$, within our framework. For small values of the
parameters $\delta_1$ and $\delta_2$, the angles $\phi$ and $\theta$
approach $\pi/4$ and $\pi/2$, respectively.}
\begin{tabular}{cc}
\includegraphics[width=7.cm]{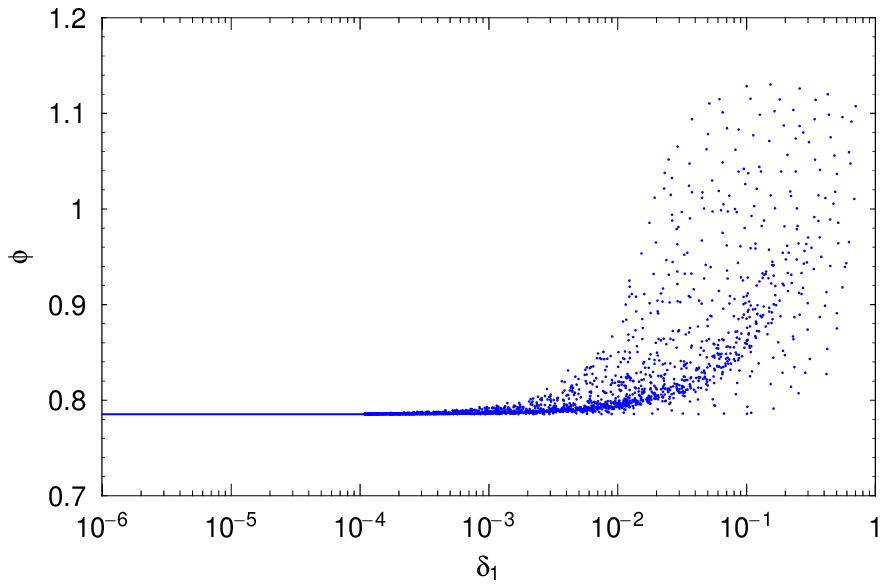}
&\includegraphics[width=7.cm]{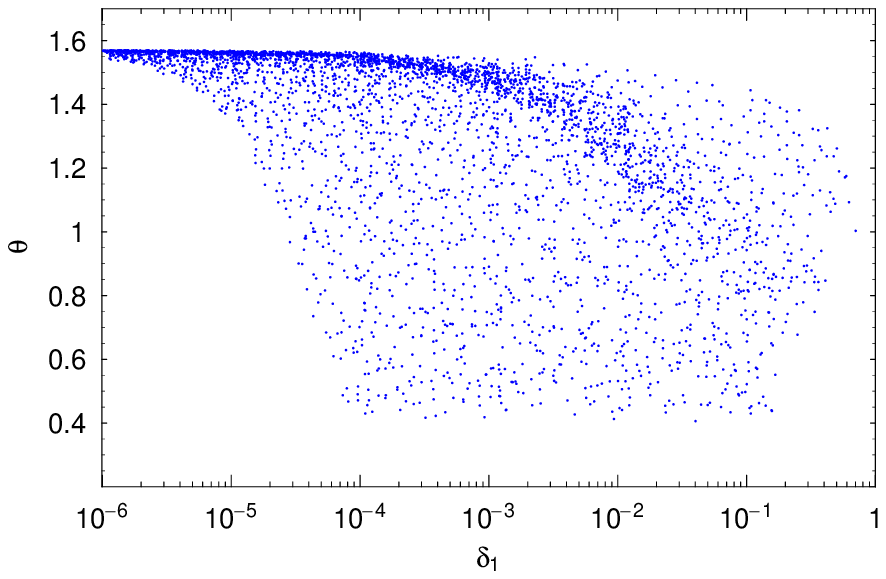}\\
\includegraphics[width=7.cm]{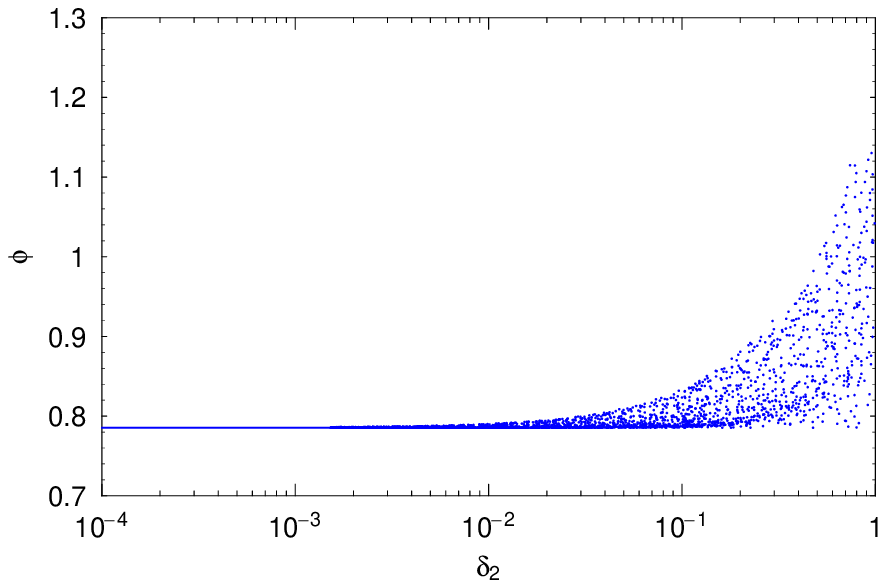}
&\includegraphics[width=7.cm]{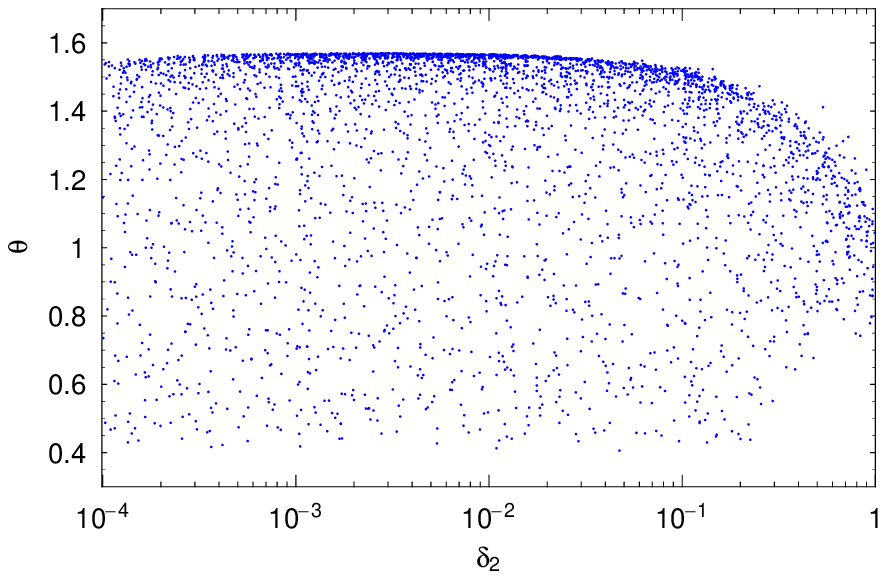}
\end{tabular}
}

\TABLE[!ht]{ \label{tab2} \caption{Numerical results for $\tan \beta
=10$. We assume right-handed neutrino degeneracy at $\LX$, which is
then lifted by radiative corrections. Two different Ans\"{a}tze were
considered for the Dirac neutrino Yukawa coupling matrix, $\Ynu =
\YU$ and $\Ynu = 3 \YU$. By slightly modifying the initial condition
$y_1/y_3 = y_u/y_t$ it is possible to reproduce the best-fit values
for all the solar and atmospheric neutrino data, as shown in the last
column of the table. The negative $CP$ parities are indicated by
$(-)$.}
\renewcommand{\tabcolsep}{1.5pc}
{\scriptsize
\begin{tabular}{lccc}
\hline  \hline \noalign{\smallskip}
& $\Ynu=\YU$     &$\Ynu=3\,\YU$  & \textbf{Best fit}  \\
\noalign{\smallskip} \hline \hline  \noalign{\smallskip}
At  $\LX=2.0 {\times} 10^{16}$~GeV  & & & \\
\noalign{\smallskip} \hline \hline  \noalign{\smallskip}
$y_u$  &$5.0 {\times} 10^{-6}$  &$5.9 {\times} 10^{-6}$    &$5.6 {\times} 10^{-6}$ \\
$y_c$  &$1.7 {\times} 10^{-3}$  &$2.1 {\times} 10^{-3}$    &$1.7 {\times} 10^{-3}$ \\
$y_t$  &0.68                  &0.80                    &0.64 \\
$y_1$  &$5.0 {\times} 10^{-6}$  &$1.8 {\times} 10^{-5}$    &$6.4 {\times} 10^{-7}$ \\
$y_2$  &$1.7 {\times} 10^{-3}$  &$6.3 {\times} 10^{-3}$    &$1.7 {\times} 10^{-3}$ \\
$y_3$  &0.68                  &2.4                     &0.64 \\
$\pi/2-\theta$  &$7.0 {\times} 10^{-3}$  &$7.0 {\times} 10^{-3}$  &$2.6 {\times} 10^{-3}$ \\
$\phi-\pi/4$    &$1.0 {\times} 10^{-5}$  &$1.0 {\times} 10^{-5}$  &$2.3 {\times} 10^{-6}$ \\
\noalign{\smallskip} \hline \hline \noalign{\smallskip}
At $\LR\,=$ &$2.6 {\times} 10^{9}$~GeV &$1.4 {\times} 10^{11}$~GeV &$6.3 {\times} 10^{8}$~GeV\\
\noalign{\smallskip}    \hline \hline \noalign{\smallskip}
$M_1$~(GeV)   & $2.4 {\times} 10^{9}$     &$9.6 {\times} 10^{10}$   &$5.8 {\times} 10^{8}$ \\
$M_2$~(GeV)   & $(-)\,2.4 {\times} 10^{9}$     &$(-)\,9.6 {\times} 10^{10}$ & $(-)\,5.8 {\times} 10^{8}$ \\
$M_3$~(GeV)   & $2.6 {\times} 10^{9}$     & $1.4 {\times} 10^{11}$   &$6.3 {\times} 10^{8}$ \\
$r_\nu$        &$6.7 {\times} 10^{-2}$    &$8.1 {\times} 10^{-2}$   &$1.4 {\times} 10^{-2}$ \\
$\tgone$       &0.34     &0.34    &0.26 \\
$\tgtwo$       &0.47     &0.57    &1.4  \\
$|U_{e3}|$       &0.18     &0.18    &0.05  \\
\noalign{\smallskip}  \hline \hline   \noalign{\smallskip}
At  $M_Z=91.2$~GeV  & & & \\
\noalign{\smallskip} \hline \hline  \noalign{\smallskip}
$K_{eff}$       & 0.31                       & 2.18                      & 0.30 \\
$m_{1}$~(eV)   &$1.6 {\times} 10^{-3}$  &$1.9 {\times} 10^{-3}$  &$1.6 {\times} 10^{-3}$ \\
$m_{2}$~(eV)   &$(-)\,1.0 {\times} 10^{-2}$  &$(-)\,1.1 {\times} 10^{-2}$  &$(-)\,6.8 {\times} 10^{-3}$ \\
$m_{3}$~(eV)   &$4.0 {\times} 10^{-2}$  &$3.9 {\times} 10^{-2}$  &$5.6 {\times} 10^{-2}$ \\
$\Delta m^2_{12}~(\text{eV}^2)$   &$1.0 {\times} 10^{-4}$
&$1.2 {\times} 10^{-4}$    &$4.3 {\times} 10^{-5}$ \\
$\Delta m^2_{23}~(\text{eV}^2)$   &$1.5 {\times} 10^{-3}$
&$1.4 {\times} 10^{-3}$    &$3.1 {\times} 10^{-3}$ \\
$r_\nu $  &$6.7 {\times} 10^{-2}$  &$8.1 {\times} 10^{-2}$ &$1.4 {\times} 10^{-2}$ \\
$\tgone$      &0.34     &0.34    &0.26 \\
$\tgtwo $      &0.47     &0.57    &1.4 \\
$|U_{e3}| $      &0.18     &0.18 &0.05  \\
$\meff$~(eV) &$2.0{\times} 10^{-12}$ &$2.4{\times} 10^{-12}$ &$9.8{\times} 10^{-15}$ \\
$m_u$~(MeV)          &1.9                &2.0      &2.2  \\
$m_c$~(GeV)          &0.68               &0.69     &0.66  \\
$m_t$~(GeV)          &174                &174      &175  \\
\noalign{\smallskip} \hline \noalign{\medskip}
Down-quark masses     &$m_d=4.61$~MeV   &$m_s=91.8$~MeV     & $m_b=2.99$~GeV \\
Charged-lepton masses & $m_e=0.49$~MeV  & $m_\mu=102.8$~MeV  & $m_\tau=1.75$~GeV \\
\noalign{\medskip} \hline\hline
\end{tabular}
}
}

\subsection{Neutrinoless double $\boldsymbol{\beta}$ decay}
\label{sec4-2}

The Majorana nature of massive neutrinos can be traced by looking for
processes where the total lepton charge $L$ changes by two units,
i.e. $\Delta L=2$. The information coming from these type of events
can be crucial not only to establish the nature of neutrinos but also
to reveal some aspects about $CP$ violation in the leptonic sector,
to which neutrino oscillations are insensitive. In particular, if
neutrinos have a Dirac signature, then the two extra Majorana phases
which appear in the neutrino mixing matrix will be absent. Moreover,
even if $CP$ invariance holds, some information about the relative
$CP$ parities and neutrino mass spectrum can be extracted. One
example of such processes is the neutrinoless double $\beta$ decay
($\beta\beta_{0\nu}\,$-\,decay) of the type
\begin{align}
(A,Z)\rightarrow(A,Z+2)+2\,e^{-}\, ,
\label{decay}
\end{align}
which can occur through the exchange of massive Majorana neutrinos
coupled to electrons. The decay rate is proportional to the
so-called ``effective Majorana mass parameter''
\begin{align}
\meff=\left|m_1\,U_{e1}^{\,2}+m_2\,U_{e2}^{\,2}+m_3\,U_{e3}^{\,2}\right|\,.
\label{meff}
\end{align}
Although no experimental evidence has been found so far for this kind
of processes, an upper bound has been obtained by the $^{76}{\rm Ge}$
Heidelberg-Moscow experiment \cite{baudis}:
\begin{align}
\meff < 0.35~\text{eV}\;[\,90\%\,{\rm C.L.}\,]\,.
\label{<m>bound}
\end{align}
If $CP$ is conserved, then Eq.~(\ref{meff}) can be written as
\begin{align}
\meff=\left|\eta_1^{CP}m_1\,U_{e1}^{\,2}+\eta_2^{CP}\,m_2\,U_{e2}^{\,2}+
\eta_3^{CP}\,m_3\,U_{e3}^{\,2}\right|\ ,\quad m_i>0\ ,
\label{mefCP}
\end{align}
where $\eta_i^{CP}={\pm} 1$ are the neutrino $CP$ parities. By examining
the invariants of the effective neutrino mass matrix given in
Eq.~(\ref{KMZfin}), one can conclude that one of the neutrinos should
have a negative $CP$ parity and the other two should have positive
$CP$ parity. This is verified for the numerical examples considered
in Table~\ref{tab2}, where the neutrino with mass $m_2$ has
$\eta_2^{CP}=-1$, while the other two have $\eta_{1,3}^{CP}=+1$. In
this case, Eq.~(\ref{mefCP}) reads
\begin{align}
\meff=\left|m_1\,U_{e1}^{\,2}-m_2\,U_{e2}^{\,2}+m_3\,U_{e3}^{\,2}\right|
\ . \label{opmef}
\end{align}
A particular feature in the above expression is the possibility of a
cancellation among the different terms \cite{wolfenstein, bilenky},
such that $\meff \simeq 0$. In fact, for all the numerical cases
considered in Table~\ref{tab2}, such a cancellation indeed occurs.
This can be easily understood if we recall that the effective
Majorana mass parameter is nothing but the absolute value of
$(\Mnu)_{11}$. Therefore, from Eq.~(\ref{KMZfin}) we have
\begin{align}
\meff \propto |\,\delta_1^2\,(c^2_{\theta}+c_{2\phi}\, s^2_{\theta})\,|\,. \label{mee}
\end{align}
Since $\theta \simeq \pi/2$ and $\phi \simeq \pi/4$, in order to
account for the low-energy neutrino data, and $\delta_1$ is small, it
is now clear from Eq.~(\ref{mee}) why $\meff$ is highly suppressed in
our framework.

\section{A simple analytical form for $\Mnu$}
\label{sec5}

In this section we shall present a simple analytical example which
can reproduce the results obtained in the last section. We have seen
that, in the context of complete degeneracy of right-handed neutrinos
and imposing proportionality between $\Ynu$ and $\YU$, the only
possible solution at low energies is the LMA. Moreover, as it is
clear from Figs. \ref{fig3} this would require $\theta \simeq \pi/2$
and $\phi \simeq \pi/4$. On the other hand, if one assumes in
Eq.~(\ref{KMZfin}), $\phi = \pi/4$ and $\theta = \pi/2$, then
\begin{align}
\kmat(M_Z)=\frac{K_{eff}}{\LR}
\left(\begin{array}{ccc}
 0 &0 & -\delta_1 \\
 0 &\delta_2^2 &0 \\
 -\delta_1 &0 &0
 \end{array}\right)\ .
 \label{pt=0}
\end{align}
In this case the neutrino masses are
 \begin{align}
 \{\,m_{1}\ ,\,m_{2}\ ,\,m_{3}\,\}= \frac{K_{eff}}{\LR}\,v^2 \sin^2 \beta\,
 \{\,\delta_1\,,\,\delta_1\,,\,\delta_2^2\,\}\ .
 \label{m1m2m3}
 \end{align}
Taking into account Eqs.~(\ref{pt=0}) and (\ref{m1m2m3}), it is
straightforward to show that, independently of the value of the
parameters $\delta_1$ and $\delta_2$, one always has $\s23=0$, which
is not phenomenologically viable.

At first sight the above result may seem contradictory, since on one
hand our numerical study clearly shows that values of $\theta \simeq
\pi/2$ and $\phi \simeq \pi/4$ are indeed compatible with the
low-energy neutrino data, but on the other hand, a simple analytical
check gives a completely opposite answer. As we shall see shortly,
the reason for this mismatching lies in the fact that small
deviations, such as $\delta \theta = \pi/2 - \theta \ll 1$ and
$\delta \phi = \phi - \pi/4 \ll 1$, and/or small perturbations from
our original assumptions, turn out to be crucial. To show that this
is indeed the case, let us relax the condition of proportionality
between $\Ynu$ and $\YU$ for the smallest Dirac neutrino Yukawa
coupling by putting
\begin{align}
 \epsilon_1=a\,\epsilon^2\ ,\quad \epsilon_2 =\epsilon
 =\frac{y_c(\LX)}{y_t(\LX)}\ ,
 \label{rel}
\end{align}
and write $\theta$ and $\phi$ as
\begin{align}
 \phi=\frac{\pi}{4}+b\,\epsilon^2\ ,\quad \theta=\frac{\pi}{2}-\epsilon\ ,
 \label{theph}
\end{align}
where $a$ and $b$ are some positive real constants. From
Eq.~(\ref{delta12}) we have then
\begin{align}
\delta_1=a\,I_{eff}\,I_{\tau}^{-1}\,\epsilon^2 \simeq a\,\epsilon^2\ , \quad \quad
\delta_2=I_{eff} \,I_{\tau}^{-1}\,\epsilon\simeq\epsilon\ .
\label{d1d2}
\end{align}

The effective neutrino mass operator (\ref{KMZfin}) will
be approximately given in this case by
\begin{align}
\kmat(M_Z)\simeq \frac{K_{eff}}{\LR}\,\epsilon^2\,\left(\begin{array}{ccc}
a^2\,\epsilon^4\,(\,1-2\,b\,) & -a\,\epsilon^2    & -a \\
-a\,\epsilon^2         &\quad 1-2\,b\,\epsilon^4 \quad & -1 \\
-a                      & -1              & 2\,b
\end{array}\right)\ .
\label{Kans1}
\end{align}
Due to the smallness of the parameter $\epsilon$ we have, to a very
good approximation,
\begin{align}
\kmat(M_Z)\simeq \frac{K_{eff}}{\LR}\,\epsilon^2\,\kmat^{\prime}\quad,\quad
\kmat^{\prime}=\left(\begin{array}{ccc}
0  & 0     & -a \\
0  & 1     & -1 \\
-a & -1    & 2\,b
\end{array}\right)\ .
\label{Kans2}
\end{align}
The invariants of the matrix $\kmat^{\prime}$ are
\begin{align}
\text{T}&=\lambda_1+\lambda_2+\lambda_3=1+2\,b \ ,\nonumber \\
\chi&=\lambda_1\,\lambda_2+\lambda_1\,\lambda_3+\lambda_2\,\lambda_3=
-1-a^2-2\,b \ , \nonumber \\
\Delta&=\lambda_1\,\lambda_2\,\lambda_3=-a^2\ ,
\end{align}
where $\lambda_i\ $ are the eigenvalues,
\begin{align}
\lambda_1&=2\,\sqrt{p}\,\cos\left(\frac{\alpha}{3}\right)+\text{T}\ ,
\nonumber \\
\lambda_{\,2,\,3}&=-2\,\sqrt{p}\,
\cos\left(\frac{\alpha}{3}{\pm}\frac{\pi}{3}\right)+\text{T} \ ,
\end{align}
with
\begin{align}
q=-\frac{2\,\text{T}^{\,3}}{27}+\frac{\text{T}\,\chi}{3}-\Delta\ ,\quad
p=\frac{\text{T}^{\,2}}{9}-\frac{\chi}{3}\ ,\quad
\cos\alpha=-\frac{q}{p^{\,3/2}}\ .
\end{align}
 The corresponding eigenvectors are given by
\begin{align}
\textbf{v}_i=\frac{1}{n_i}\left[\,a\,(\,1-\lambda_i\,)\,,\,-\lambda_i\,,\,
\lambda_i\,(\,\lambda_i-1\,)\,\right]^{\,T} \ ,
\label{eig}
\end{align}
where $n_i$ are the normalization coefficients.

In Fig.~\ref{fig4} we present the region in the $(a,b)$-plane, which
is compatible with the low-energy solar and atmospheric neutrino data
at 90\% and 99\% C.L.\ for the LMA solution (cf. Table~\ref{tab1}).
We see that no fine-tuning is required in order to reproduce the
data. To obtain the best-fit point (black dot in the figure), we
assume at the unification scale $\LX$ the same values for the charm
and top Yukawa couplings as in the last column of Table \ref{tab2},
i.e. $y_c(\LX)=1.7 {\times} 10^{-3}$ and $y_t(\LX)=0.64\ $. Taking then as
input parameters
\begin{align}
\epsilon=\frac{y_c(\LX)}{y_t(\LX)} = 2.65 {\times} 10^{-3}\ ,
\quad a=0.15 \ ,\quad b=0.34
\ ,\quad \LR=6.3 {\times} 10^8~\text{GeV}\ ,
\label{best1}
\end{align}
the following values are obtained at the electroweak scale for $\tan
\beta=10\ $:
\begin{align}
\begin{tabular}{lll}
$K_{eff}=0.3$\ ,  & $\delta_1=1.0 {\times} 10^{-6}$\ ,  &
 $\delta_2=2.7 {\times} 10^{-3}$\ , \medskip \\
 $m_{1}=1.6 {\times} 10^{-3}$~eV\ ,
 &$m_{2}=6.7 {\times} 10^{-3}$~eV\ ,
& $m_{3}=5.6 {\times} 10^{-2}$~eV\ , \medskip \\
 $\Delta m^2_{12}=4.2 {\times} 10^{-5}~\text{eV}^2$\ ,  &
 $\Delta m^2_{23}=3.1 {\times} 10^{-3}~\text{eV}^2$\ ,
& $r_\nu=1.4 {\times} 10^{-2}$\ , \medskip \\
 $\tgone=0.26$\ ,  & $\tgtwo=1.4$\ ,  & $|U_{e3}|=0.05$\ , \medskip \\
 $\meff=9.3 {\times} 10^{-15}$  eV\ , & &
\end{tabular}
\label{best2}
\end{align}
which correspond to the best-fit values in the neutrino mixing plane
for the atmospheric neutrinos \cite{gonzalez} and the LMA solar
neutrino solution \cite{bahcall}.

\FIGURE[!ht]{ \label{fig4} \caption{Allowed region for the parameters
$a$ and $b$ of the triangular texture (\ref{Kans2}) for $\kmat(M_Z)$,
in the case of the LMA MSW solution and considering the neutrino data
at 90\% and 99\% C.L.\ summarized in Table~\ref{tab1}. The black dot
corresponds to the point which leads to the best-fit values of the
solar and atmospheric data as shown in
Eqs.~(\ref{best1})-(\ref{best2}).}
$$\includegraphics[width=9cm]{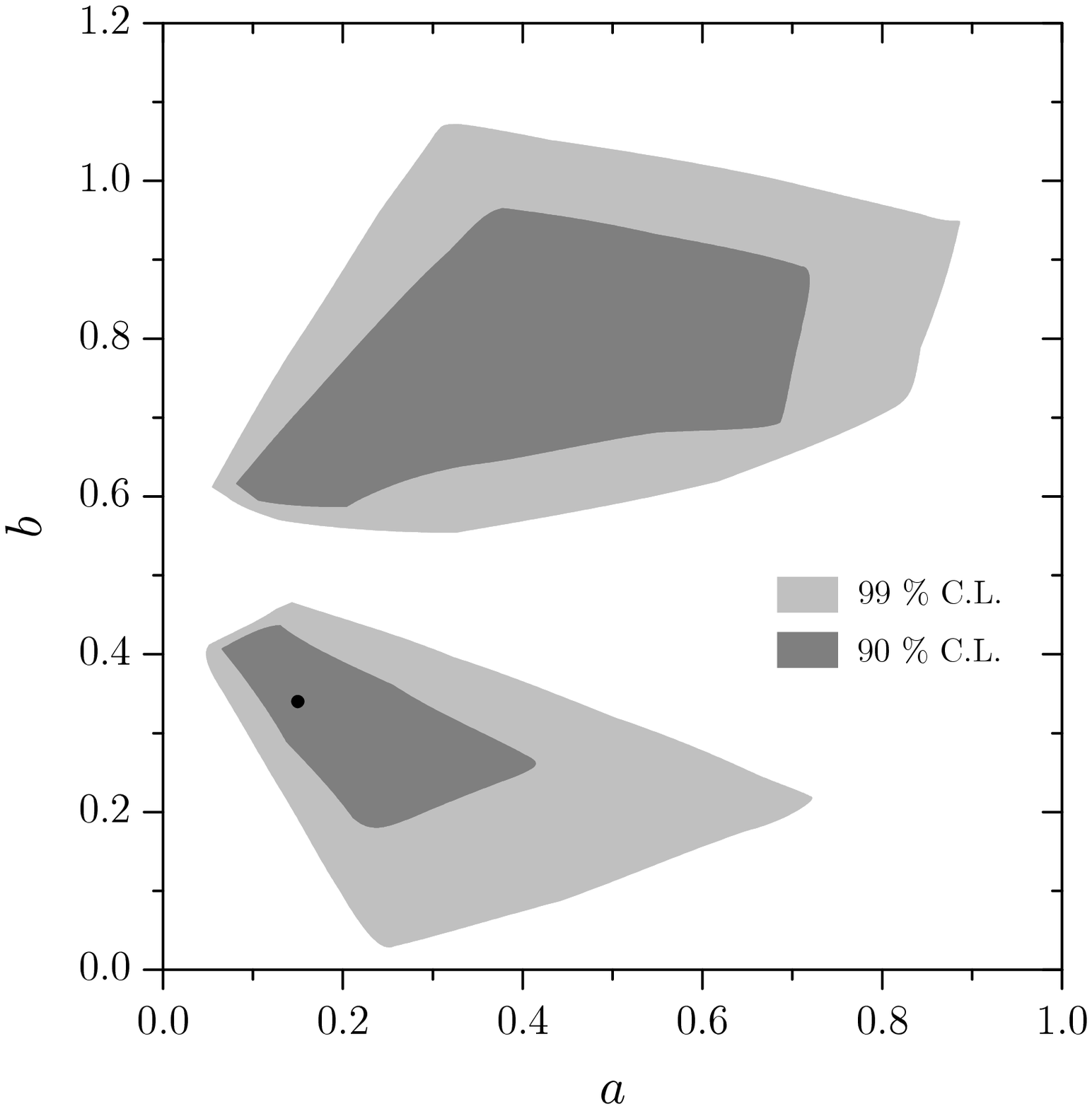}$$
}

\section{Conclusions}
\label{sec6}

In this paper we have considered the possibility of having an exactly
degenerate spectrum for heavy right-handed Majorana neutrinos at the
grand unification scale. Our theoretical framework was based on the
minimal supersymmetric standard model with unbroken $R$ parity and
extended with three heavy Majorana neutrino fields. We assumed no
Dirac-type leptonic mixing at GUT scale, and thus the right-handed
neutrino sector constitutes the only source of lepton flavour
violation in this context. We have then studied the renormalization
group effects, which include the neutrino threshold effects, from
high to low energies.

Taking into account the allowed range for the quark mass ratios at
GUT scale and the effective neutrino mass matrix obtained through the
seesaw mechanism, we have investigated which of the possible
solutions to the solar neutrino problem can be accommodated in our
framework. Inspired by GUT-motivated relations among the quark,
charged-lepton and Dirac neutrino mass matrices, and based on the
latest global analyses of the solar and atmospheric data at 99\%
C.L., we have then concluded that the only solar solutions compatible
with the experimental data are the LOW and LMA solutions. In fact, at
90\% C.L. only the latter is allowed. Even in this case, the obtained
values for the neutrino oscillation parameters are far from the
best-fit points. In this sense, right-handed neutrino degeneracy
embedded in GUT scenarios seems to be disfavoured by the low-energy
neutrino data. We emphasize however that, by slightly relaxing the
imposed relations among the Yukawa coupling matrices at high scale,
one can easily reach the best-fit values. Our assumption of
degeneracy is in fact a very restrictive one, since the right-handed
Majorana neutrino matrix $\MR$ is parametrized in terms of only two
angles and an overall scale $\LR$. On the other hand, a fully or
partially hierarchical mass spectrum for right-handed neutrinos can
be easily achieved with a much less constrained structure, and one
could therefore expect that such patterns are easier to be reconciled
with the low-energy neutrino data.

Finally, we are also aware of possible contributions to lepton
flavour violation processes like $\mu \rightarrow e \gamma\,$.  As it
has been recently pointed out by Casas and Ibarra \cite{casas}, in
supersymmetric theories with seesaw-like neutrino masses the
prediction for the branching ratio of this process is, in general,
larger than the experimental upper bound. In particular, scenarios
such as the one considered here, where top-neutrino unification
occurs, are severely constrained unless the leptonic Yukawa coupling
matrices satisfy very specific requirements.

\acknowledgments

We thank E.Kh.\,Akhmedov for a careful reading of the manuscript and
useful comments. We are also grateful to D.F.~Carvalho for bringing to
our attention some aspects of lepton flavour violation in $\mu \rightarrow e
\gamma\,$. The work of R.G.F. and F.R.J. has been supported by {\em Funda\c{c}\~{a}o
para a Ci\^{e}ncia e a Tecnologia} under the grants SFRH/BPD/1549/2000
and PRAXISXXI/BD/18219/98, respectively.


\appendix

\section{Renormalization group equations}

In this Appendix we present the relevant one-loop renormalization
group equations for the MSSM extended with three right-handed
Majorana neutrinos.

Between $\LX$ and $\LR$ the evolution of  the up-quark, $\YU\,$,
down-quark, $\YD\,$, charged-lepton, $\Yl\,$, and Dirac neutrino, $\Ynu\,$,
Yukawa coupling matrices is given by

\begin{align}
16\,\pi^2\frac{d\YU}{dt}&=\left[\,(\,T_U-G_U\,)+(\,3\,\YU\YU^\dagger
+\YD\YD^\dagger\,)\,\right]\YU\ ,
\label{YUrge}\\
16\,\pi^2\frac{d\YD}{dt}&=\left[\,(\,T_D-G_D\,)+(\,3\,\YD\YD^\dagger
+\YU\YU^\dagger\,)\,\right]\YD\ ,
\label{YDrge}\\
16\,\pi^2\frac{d\Yl}{dt}&=\left[\,(\,T_{\ell}-G_{\ell}\,)+(\,3\,\Yl\Yl^\dagger
+\Ynu\Ynu^\dagger\,)\,\right]\Yl\ ,
\label{Ylrge}\\
16\,\pi^2\frac{d\Ynu}{dt}&=\left[\,(\,T_{\nu}-G_{\nu}\,)+(\,3\,\Ynu\Ynu^\dagger
+\Yl\Yl^\dagger\,)\,\right]\Ynu\ , \label{Ynurge}
\end{align}
where $t=\ln (\mu/M_Z)$, $\mu$ is the renormalization scale. The
quantities $T_i$ and $G_i$ are defined as
\begin{align}
&T_U=\text{Tr}\,(\,3\,\YU\YU^\dagger+\Ynu\Ynu^\dagger\,)\ ,\
G_U=\frac{16}{3}\,g_3^2+3\,g_2^2+\frac{13}{15}\,g_1^2\ ,\\
&T_D=\text{Tr}\,(\,3\,\YD\YD^\dagger+\Yl\Yl^\dagger\,)\ ,\
G_D=\frac{16}{3}\,g_3^2+3\,g_2^2+\frac{7}{15}\,g_1^2\ ,\\
&T_{\ell}=\text{Tr}\,(\,3\,\YD\YD^\dagger+\Yl\Yl^\dagger\,)\ ,\
G_{\ell}=3\,g_2^2+\frac{9}{15}\,g_1^2\ ,\\
&T_{\nu}=\text{Tr}\,(\,3\,\YU\YU^\dagger+\Ynu\Ynu^\dagger\,)\ ,\
G_{\nu}=3\,g_2^2+\frac{3}{5}\,g_1^2\ ,
\end{align}
where $g_3$, $g_2$, and $g_1$ are the $SU(3)$, $SU(2)$ and $U(1)$
gauge couplings, which satisfy the RGE
\begin{align}
16\,\pi^2\frac{d\,g_i}{dt}=b_i\,g_i^3\quad,\quad
b_i=\left(33/5,1,-3\right)\ .
\label{grge}
\end{align}

The evolution of the right-handed neutrino mass matrix, $\MR\,$, is
given by
\begin{align}
8\,\pi^2\frac{d\MR}{dt}=\MR\,(\Ynu\Ynu^\dagger)^T\,+\,(\Ynu\Ynu^\dagger)\,\MR\ .
\label{MRrge}
\end{align}

Below $\LR\,$, the renormalization group equations for the up-quark
and charged-lepton Yukawa couplings get modified since after
decoupling of the right-handed neutrino states, the contribution of
$\Ynu$ should vanish. In this case the RGE for $\YU$ and $\Yl$ read
\begin{align}
16\,\pi^2\frac{d\YU}{dt}&=\left[\,(\,T_U^{\,\prime}-G_U\,)+(\,3\,\YU\YU^\dagger
+\YD\YD^\dagger\,)\,\right]\YU\,,
\label{YUrge2}\\
16\,\pi^2\frac{d\Yl}{dt}&=\left[\,(\,T_{\ell}-G_{\ell}\,)+3\,\Yl\Yl^\dagger
\,\right]\Yl\ ,
\label{Ylrge2}
\end{align}
with
\begin{align}
T_U^{\,\prime}=3\,\text{Tr}\,(\YU\YU^\dagger)\,.
\end{align}

Finally, the evolution of the effective neutrino mass matrix $\kmat$
from $\LR$ to $M_Z$ is given by
\begin{align}
16\,\pi^2\frac{d\kmat}{dt}=(\,T_{\kappa}-G_{\kappa}\,)\kmat
+(\,\Yl\Yl^\dagger\,)\,\kmat+\kmat\,(\,\Yl\Yl^\dagger\,)^T\ ,
\label{krge}
\end{align}
where
\begin{align}
T_\kappa=6\,\text{Tr}\,(\,\YU\YU^\dagger\,)\quad,\quad
G_\kappa=6\,g_2^2+\frac{6}{5}\,g_1^2\ .
\end{align}


\begin{thebibliography}{999}

\bibitem{SK1} Y. Fukuda \emph{et al.}, Super-Kamiokande
Collaboration,
\plb{433}{1998}{9}; \plb{436}{1998}{33};
\plb{467}{1999}{185}; \prl{82}{1999}{2644};
\prl{85}{2000}{3999}.

\bibitem{CHOOZ} M. Apollonio \emph{et al.}, CHOOZ Collaboration,
\plb{420}{1998}{397}; \plb{466}{1999}{415}.

\bibitem{SK2} Y. Fukuda \emph{et al.}, Super-Kamiokande
Collaboration, \prl{81}{1998}{1158}; \prl{81}{1998}{4279(E)};
\prl{82}{1999}{1810}; \prl{82}{1999}{2430}; \prl{86}{2001}{5651};
\prl{86}{2001}{5656}.

\bibitem{homestake} B. T. Cleveland \emph{et al.}, {\it Astrophys. J.}
\textbf{496} (1998) 505; R. Davis, \ppnp{32}{1994}{13}.

\bibitem{sage} J.N. Abdurashitov \emph{et al.}, SAGE Collaboration,
\prc{60}{1999}{055801}.

\bibitem{gallex} W. Hampel \emph{et al.}, GALLEX Collaboration,
\plb{447}{1999}{127}.

\bibitem{MSW} L. Wolfenstein, \prd{17}{1978}{2369}; S.P. Mikheyev,
A. Yu Smirnov, \sjnp{42}{1986}{913}.

\bibitem{seesaw} M. Gell-Mann, P. Ramond and R. Slansky, in
\emph{Supergravity}, eds. P. van Nieuwenhuizen and D.Z. Freedman
(North Holland, Amsterdam, 1979), p. 315; T. Yanagida, in Proc. of
the Workshop on the Unified Theory and Baryon Number in the Universe,
KEK report 79-18 (1979), p. 95; R.N. Mohapatra and G. Senjanovic,
\prl{44}{1980}{912}.

\bibitem{so10} For a recent review on grand unified theories see
\emph{e.g.} R.N. Mohapatra, \emph{Lectures at ICTP Summer School
in Particle Physics, Trieste, Italy, 7 June - 9 July 1999},
\hepph{9911272}, and references therein.

\bibitem{georgi} H. Georgi and C. Jarlskog, \plb{86}{1979}{297}.

\bibitem{harvey} J. Harvey, P. Ramond and D. Reiss, \plb{92}{1980}{309};
X.G. He and S. Meljanac, \prd{41}{1990}{1620}; S. Dimopoulos, L. Hall and
S. Raby, \prl{68}{1992}{1984}; K.S. Babu and R.N. Mohapatra,
\prl{74}{1995}{2418}.

\bibitem{smirnov} A.Yu. Smirnov, \prd{48}{1993}{3264}.

\bibitem{lola} G.K. Leontaris, S. Lola, C. Scheich and J. Vergados,
\prd{53}{1996}{6381}; S. Lola and J. Vergados, \ppnp{40}{1998}{71};
J. Ellis, G.K. Leontaris, S. Lola and D.V. Nanopoulos,
\epjc{9}{1999}{389}.

\bibitem{binetruy} P. Binetruy, S. Lavignac, S. Petcov and P. Ramond,
\npb{496}{1997}{3}.

\bibitem{akhmedov} E. Kh. Akhmedov, G.C. Branco and M.N. Rebelo,
\plb{478}{2000}{215}.

\bibitem{joaquim} E. Kh. Akhmedov, G.C. Branco, F.R. Joaquim and
J.I. Silva-Marcos, \plb{498}{2001}{237}.

\bibitem{altarelli} G. Altarelli and F. Feruglio,
\plb{451}{1999}{388}; C.H. Albright and S.M. Barr,
\prd{64}{2001}{073010}.

\bibitem{ellis} J. Ellis and S. Lola, \plb{458}{1999}{310};
J. A. Casas, J. R.  Espinosa, A. Ibarra and I. Navarro,
\npb{569}{2000}{82}.

\bibitem{dighe} A.S. Dighe and A.S. Joshipura, \hepph{0010079}.

\bibitem{chun} E.J. Chun, \plb{505}{2001}{155}.

\bibitem{RGE} P.H. Chankowski and Z. Pluciennik, \plb{316}{1993}{312};
K. Babu, C. N. Leung and J. Pantaleone, \plb{319}{1993}{191};
M. Tanimoto, \plb{360}{1995}{41}; J. Ellis and S. Lola, \plb{458}{1999}{310};
N. Haba {\it et al.}, \ptp{103}{2000}{367}; \epjc{10}{1999}{177};
\epjc{14}{2000}{347};
J. A. Casas, J. R.  Espinosa, A. Ibarra and I. Navarro,
\npb{556}{1999}{3}; \jhep{9909}{1999}{015};
\npb{573}{2000}{652}; E. Ma, J. Phys. {\bf G25} (1999) 97;
R. Barbieri, G.G. Ross and A. Strumia, \jhep{9910}{1999}{020};
P.H. Chankowski, W. Krolikowski and S. Pokorski, \plb{473}{2000}{109};
S.F. King and N.N. Singh, \npb{591}{2000}{3}.

\bibitem{branco} G. C. Branco, M. N. Rebelo and J. I. Silva-Marcos;
\prl{82}{1999}{683}.

\bibitem{weinberg} S. Weinberg, \prl{43}{1979}{1566}.

\bibitem{babu} K.S. Babu, C.N. Leung and J. Pantaleone, \plb{319}{1993}{191}.

\bibitem{PDG} D.E. Groom \emph{et al.}, Particle Data Group,
\epjc{15}{2000}{1}.

\bibitem{bilenky} S.M. Bilenky, S. Pascoli and S.T. Petcov,
\prd{64}{2001}{053010}.

\bibitem{gonzalez} M.C. Gonzalez-Garcia, M. Maltoni, C. Pena-Garay and
J.W.F. Valle, \prd{63}{2001}{033005}.

\bibitem{bahcall} J.N. Bahcall, P.I. Krastev and A.Yu. Smirnov,
\jhep{0105}{2001}{015}.

\bibitem{koide} H. Fusaoka and Y. Koide, \prd{57}{1998}{3986}.

\bibitem{baudis} L. Baudis \emph{et al.}, \prl{83}{1999}{41}.

\bibitem{wolfenstein} L.Wolfenstein, \plb{107}{1981}{77}.

\bibitem{casas} J.A. Casas and A. Ibarra, \hepph{0103065}.

\end{thebibliography}
\end{document}